\documentclass[aps,prd,twocolumn,floatfix,amsmath,nofootinbib,amssymb,preprintnumbers]{revtex4-2}
\usepackage{graphicx,amsmath,mathtools,color,dcolumn,booktabs,diagbox,hyperref,placeins}
\usepackage{txfonts}
\usepackage{slashed}
\usepackage{epstopdf}
\usepackage{multirow}
\usepackage{bm}
\usepackage{braket}
\usepackage{pifont}
\usepackage{adjustbox}
\usepackage{rotating}
\usepackage{verbatim}
\usepackage{xcolor}
\usepackage{appendix}
\begin{document}
\hypersetup{hidelinks}

\title{Exploring charmonia $\eta_c(6S,7S)$ in the $\Lambda_c^+\bar{\Lambda}_c^-$ invariant mass spectrum of $B^0 \to K_S^0\Lambda_c^+\bar{\Lambda}_c^-$}

\author{Cheng-Xi Liu$^{1,2,3,4}$}\email{liuchx2025@lzu.edu.cn}
\author{Jun-Zhang Wang$^5$}\email{wangjzh@cqu.edu.cn}
\author{Xiang Liu$^{1,2,3,4}$}
\email{xiangliu@lzu.edu.cn}

\affiliation{
$^1$School of Physical Science and Technology, Lanzhou University, Lanzhou 730000, China\\
$^2$Lanzhou Center for Theoretical Physics, Key Laboratory of Theoretical Physics of Gansu Province, Key Laboratory of Quantum Theory and Applications of MoE, Gansu Provincial Research Center for Basic Disciplines of Quantum Physics, Lanzhou University, Lanzhou 730000, China\\
$^3$MoE Frontiers Science Center for Rare Isotopes, Lanzhou University, Lanzhou 730000, China\\
$^4$Research Center for Hadron and CSR Physics, Lanzhou University and Institute of Modern Physics of CAS, Lanzhou 730000, China\\
$^5$Department of Physics and Chongqing Key Laboratory for Strongly Coupled Physics, Chongqing University, Chongqing 401331, China}

\begin{abstract}
We analyze the $B^0 \to K_S^0 \Lambda_c^+ \bar{\Lambda}_c^-$ decay recently observed by the LHCb Collaboration. The $\Lambda_c^+\bar{\Lambda}_c^-$ invariant-mass spectrum exhibits an accumulation near 4.63~GeV, which we investigate together with the accompanying $\Lambda_c^+ K_S^0$ spectrum constrained by the known $\Xi_c(2923)$ and $\Xi_c(2939)$ states. Our amplitude model includes a nonresonant term, the two established $\Xi_c$ contributions, and high-lying charmonium states in the $\Lambda_c^+\bar{\Lambda}_c^-$ channel. Guided by screened Godfrey--Isgur and quark-pair-creation calculations, we test the 4.63~GeV enhancement with a fixed $\eta_c(6S)$ amplitude at $(M_1,\Gamma_1)=(4629.00,24.20)$~MeV, {which improves the description of the 4.63~GeV region}, and examine a fixed $\eta_c(7S)$ contribution at $(M_2,\Gamma_2)=(4718.84, 22.84)$~MeV, which improves the higher-mass region of the spectrum. We also evaluate the $\eta_c(nS) \to \Lambda_c^+ \bar{\Lambda}_c^-$ baryonic decays via a hadron-loop mechanism and estimate the $B^0 \to K_S^0 \eta_c(nS)$ production in naive factorization. Within the narrow width approximation, the fit yields a $\mathcal B(B^0\to K_S^0\eta_c(6S)) \times \mathcal B(\eta_c(6S)\to\Lambda_c^+\bar{\Lambda}_c^-)$ of about $6.2 \times 10^{-6}$, while the factorization and hadron-loop calculations give rates roughly one to three orders of magnitude lower than the above fitted ratio. These results highlight the need for more precise data and a full amplitude analysis to determine the line shapes and clarify the nature of the higher-mass contribution.

\end{abstract}

\maketitle

\section{Introduction}

Recently, the LHCb Collaboration reported the first observation of the decay $B^0 \to \Lambda_c^+ \bar{\Lambda}_c^- K_S^0$ using $pp$ collision data at $\sqrt{s}=13~\text{TeV}$, corresponding to an integrated luminosity of $5.4~\text{fb}^{-1}$~\cite{LHCb:2604.15040}. The branching fraction is measured to be $\mathcal{B}(B^0 \to \Lambda_c^+ \bar{\Lambda}_c^- K_S^0) \approx 2.6 \times 10^{-4}$. More importantly, LHCb found evidence for two $\Xi_c^+$ states, $\Xi_c(2923)^+$ and $\Xi_c(2939)^+$, in the $\Lambda_c^+ K_S^0$ invariant mass spectrum. Prior to this observation, the decays $\bar{B}^0 \to \bar{K}^0 \Lambda_c^+ \bar{\Lambda}_c^-$ and $B^- \to \Lambda_c^+ \bar{\Lambda}_c^- K^-$ were investigated by the Belle~\cite{Belle:2018} and LHCb~\cite{LHCb:2023} collaborations, respectively.

Upon careful inspection of the $\Lambda_c^+ \bar{\Lambda}_c^-$ invariant mass spectrum of $B^0 \to \Lambda_c^+ \bar{\Lambda}_c^- K_S^0$ recently reported by LHCb, one may notice an accumulation of events around $4.63~\text{GeV}$. {The corresponding spectra measured in $\bar{B}^0 \to \bar{K}^0 \Lambda_c^+ \bar{\Lambda}_c^-$~\cite{Belle:2018} and $B^- \to \Lambda_c^+ \bar{\Lambda}_c^- K^-$~\cite{LHCb:2023} also show event accumulations around $4.63~\text{GeV}$, although neither analysis established a statistically significant $\Lambda_c^+\bar{\Lambda}_c^-$ structure in this region.} This observation naturally motivates our interest in investigating the nature of this enhancement.

In fact, a charmonium-like structure, $Y(4630)$, was previously reported in $e^+e^- \to \Lambda_c^+ \bar{\Lambda}_c^-$ by Belle~\cite{Pakhlova:2008}. As discussed in Ref.~\cite{Wang:2020yaa}, $Y(4630)$ can be interpreted as a good candidate for a high-lying charmonium state within an $S$-$D$ mixing scheme of vector charmonium. This naturally suggests that studies of the $\Lambda_c^+ \bar{\Lambda}_c^-$ invariant mass spectrum can provide valuable insights for constructing the charmonium family. Recently, the unquenched effects, which incorporate coupled-channel contributions such as hadronic loops, have been shown to be essential for a unified description of the charmonium spectrum and the exotic states above open-charm thresholds~\cite{bai:2026atm}. The enhancements observed in $B$ decays and $e^+e^-$ annihilation may have different origins because the two production processes allow different $J^{PC}$ quantum numbers~\cite{Wang:2025dur}.

Returning to the newly observed $B^0 \to \Lambda_c^+ \bar{\Lambda}_c^- K_S^0$ decay, we carry out a combined analysis of the measured $\Lambda_c^+\bar{\Lambda}_c^-$ and $\Lambda_c^+K_S^0$ invariant mass spectra. In addition to the known $\Xi_c(2923)^+$ and $\Xi_c(2939)^+$ contributions to the $\Lambda_c^+K_S^0$ invariant mass as observed by LHCb~\cite{LHCb:2604.15040}, we test $\eta_c(6S)$ and $\eta_c(7S)$ amplitudes in the $\Lambda_c^+\bar{\Lambda}_c^-$ invariant mass spectrum {using fixed representative masses and widths selected from the ranges obtained in our screened-GI and QPC calculations. These calculations follow the spectroscopic framework of Refs.~\cite{Wang:2019,Wang:2020yaa}.} Here, $\eta_c(6S)$ and $\eta_c(7S)$ are suitable as intermediate states of $B^0 \to \Lambda_c^+ \bar{\Lambda}_c^- K_S^0$, since their production and baryonic decay proceed in an $S$ wave. The $\eta_c(6S)$ term gives the main improvement near 4.63~GeV, and the $\eta_c(7S)$ term probes the higher-mass region. Future LHCb and Belle~II data can resolve the corresponding line shapes.

In this work, we further calculate the open-charm and $\Lambda_c^+\bar{\Lambda}_c^-$ decay rates of the two discussed $\eta_c$ charmonia, and roughly estimate $B^0\to K_S^0\eta_c(6S,7S)$ production in naive factorization. Together with the fitted fraction, these calculations provide a first comparison of the full $\eta_c(6S)$ production and decay chain. These results provide a phenomenological reference for future searches for high-lying $\eta_c$ states in baryonic $B$-meson decays.

The remainder of this paper is organized as follows. In Sec.~\ref{sec:fit}, we construct the coherent amplitude model, perform the baseline fit, test the fixed $\eta_c(6S)$ and $\eta_c(7S)$ amplitudes, and compare them with the charmonium spectrum. In Sec.~\ref{sec:phenomenology}, we calculate the baryonic decays of the $\eta_c(6S)$ and $\eta_c(7S)$ {states considered here}, estimate their weak production in naive factorization, and compare these estimates with the fitted fraction. Section~\ref{sec:summary} contains a summary and outlook. 
\section{Line-shape analysis of the $\Lambda_c^+\bar{\Lambda}_c^-$ and $\Lambda_c^+K_S^0$ invariant mass spectra of $B^0\to\Lambda_c^+\bar\Lambda_c^-K_S^0$}
\label{sec:fit}

The $\Lambda_c^+\bar{\Lambda}_c^-$ invariant-mass spectrum measured in $B^0\to\Lambda_c^+\bar{\Lambda}_c^-K_S^0$ shows an accumulation of events near 4.63~GeV. This region is well above the $\Lambda_c^+\bar{\Lambda}_c^-$ threshold and may receive contributions from highly excited hidden-charm states. We study the enhancement together with the accompanying $\Lambda_c^+K_S^0$ spectrum, where the $\Xi_c(2923)$ and $\Xi_c(2939)$ signals constrain the three-body amplitude. Starting from a nonresonant term and the two known $\Xi_c$ states, we construct the coherent amplitude model and perform a baseline fit to the two spectra. Guided by the quantum numbers of the fitted enhancement, we then turn to the $\eta_c(6S)$ and $\eta_c(7S)$ charmonium states and their predicted widths. Finally, we use these spectroscopically constrained masses and widths as the resonant terms in the amplitude and fit the data, finding that they accommodate the observed structures.

\subsection{Amplitude model and baseline fit}
\label{sec:coherent-model}

For bin $i$ of projection $\alpha$, the expected signal contribution is obtained by integrating the intensity $\mathcal I=|\mathcal M|^2$ over the other independent invariant mass,
\begin{equation}
	S_{i,\alpha}(\boldsymbol\theta)=\int_{s_{\alpha,i}^{\rm low}}^{s_{\alpha,i}^{\rm high}} ds_\alpha \int_{s_\beta^{\rm min}(s_\alpha)}^{s_\beta^{\rm max}(s_\alpha)} ds_\beta\, \varepsilon(s_\alpha,s_\beta)\mathcal I(s_\alpha,s_\beta;\boldsymbol\theta).
	\label{eq:fit-bin-integral}
\end{equation}
Here, $i$ labels an experimental mass bin, and $\alpha,\beta\in\{12,13,23\}$ with $\alpha\neq\beta$ label the two independent invariant-mass variables. The outer integration limits are the squared mass edges of bin $i$, $s_{\alpha,i}^{\rm low/high}=(m_{\alpha,i}^{\rm low/high})^2$, while the inner limits follow from three-body kinematics at fixed $s_\alpha$. The $m_{ij}$ is defined with
\begin{equation}
 m_{12}=m(\Lambda_c^+\bar{\Lambda}_c^-),\ 
 m_{13}=m(\Lambda_c^+K_S^0),\ 
 m_{23}=m(\bar{\Lambda}_c^-K_S^0).
 \label{eq:fit-invariant-masses}
\end{equation}  
The integration covers the physical three-body region. For the $\Lambda_c K_S^0$ spectrum, the two combinations are included as $S_{i,13}+S_{i,23}$, whereas the $\Lambda_c^+\bar{\Lambda}_c^-$ spectrum is given by $S_{i,12}$. The published spectra contain no efficiency map, and we therefore take $\varepsilon(s_{12},s_{13})=1$.

All resonant terms are represented by the relativistic Breit--Wigner function adopted in Refs.~\cite{LHCb:2019, LHCb:2604.15040},
\begin{equation}
 {\rm RBW}_R(s)=\frac{M_R\Gamma_R}  {M_R^2-s-iM_R\Gamma_R(s)}.
 \label{eq:fit-rbw}
\end{equation}
The running width is
\begin{equation}
 \Gamma_R(s)=\Gamma_R \left(\frac{q(s)}{q_R}\right)^{2L+1} \left(\frac{M_R}{\sqrt{s}}\right) \left[\frac{B_L(q(s))}{B_L(q_R)}\right]^2 .
 \label{eq:fit-running-width}
\end{equation}
Here, $q(s)$ denotes the relative momentum of the two decay products in the resonance rest frame, and $q_R\equiv q(M_R^2)$ denotes the corresponding value at the resonance pole. The quantity $B_L$ is the Blatt--Weisskopf factor. We use $L=2$ for the two established $\Xi_c$ contributions.  For the $\eta_c(nS)$ terms we use the lowest-partial-wave form, $L=0$, as a phenomenological parametrization.  Their possible quantum numbers are tested later through the spectroscopic comparison.

The signal amplitude contains a nonresonant contribution and the two established $\Xi_c$ line shapes.  The nonresonant term is represented by a constant amplitude $A_{\rm NR}$ over the physical region; its mass projections are shaped by phase space. The coherent signal form used below is
\begin{equation}
 \begin{aligned}
  \mathcal M_{\Xi_c}(s_{13}) &=A_{\rm NR} +C_{\Xi_c^{(1)}}{\rm RBW}_{\Xi_c^{(1)}}(s_{13})\\
   &\quad+C_{\Xi_c^{(2)}}{\rm RBW}_{\Xi_c^{(2)}}(s_{13}),\\
 \mathcal I_0^{\rm coh}&=|\mathcal M_{\Xi_c}(s_{13})|^2.
 \end{aligned}
 \label{eq:fit-baseline-intensities}
\end{equation}
The $\Xi_c$ masses and widths are fixed to the LHCb values~\cite{LHCb:2604.15040}, whereas their complex coefficients $C_{\Xi_c^{(1)}}$ and $C_{\Xi_c^{(2)}}$ are fitted. The fixed experimental background is added to the signal after the phase-space integration. The expected bin contents are
\begin{equation}
{\begin{aligned}
 \mu_{i,1}(\boldsymbol\theta)
 &=\nu_1\left[B_{i,1}+S_{i,13}(\boldsymbol\theta)+S_{i,23}(\boldsymbol\theta)\right],\\
 \mu_{i,2}(\boldsymbol\theta)&=B_{i,2}+S_{i,12}(\boldsymbol\theta).
\end{aligned}}
	\label{eq:fit-total-bin}
\end{equation}
Here, label 1 denotes the published $m(\Lambda_c K_S^0)$ spectrum, which contains both charge-conjugate combinations, and label 2 denotes the $m(\Lambda_c^+\bar\Lambda_c^-)$ spectrum. $B_{i,k}$ is the background shape read from the LHCb spectrum and $S_{i,\alpha}$ is given by Eq.~\eqref{eq:fit-bin-integral}. One overall normalization is absorbed into the amplitude coefficients, while $\nu_1$ accounts for the relative normalization of the two published projections and is fixed by the total $m(\Lambda_c K_S^0)$ yield. The background does not interfere with the signal.

For each projection, the fit quality is evaluated using the published asymmetric uncertainty on the side of the fitted value,
\begin{equation}
 {\chi_k^2=\sum_i \frac{[n_{i,k}-\mu_{i,k}(\boldsymbol\theta)]^2} {\sigma_{i,k}^2},\qquad k=1,2.}
 \label{eq:fit-projection-chi2}
\end{equation}
Thus, $\chi_1^2$ and $\chi_2^2$ measure the fit quality of the $m(\Lambda_c K_S^0)$ and $m(\Lambda_c^+\bar\Lambda_c^-)$ spectra, respectively. Here, $\sigma_{i,k}$ is chosen from the upper or lower quoted uncertainty according to whether the fitted value lies above or below the central bin content. 

In our work, the quoted $\chi_k^2$ values are not divided by the number of degrees of freedom. The experimental error bars are large compared with the differences between the fitted curves and the data, so even the baseline gives small $\chi_k^2/N_{\rm dof}$ values. Though adding the charmonium amplitudes improves the line-shape description, it also reduce $\chi_k^2/N_{\rm dof}$ values. When the $\chi_k^2/N_{\rm dof}$ is already below 1, a further decrease does not by itself indicate a worse fit. We therefore use the decrease in $\chi_k^2$ to describe the improvement in each spectrum.

\begin{figure}[htbp]
\centering
\includegraphics[width=\columnwidth]{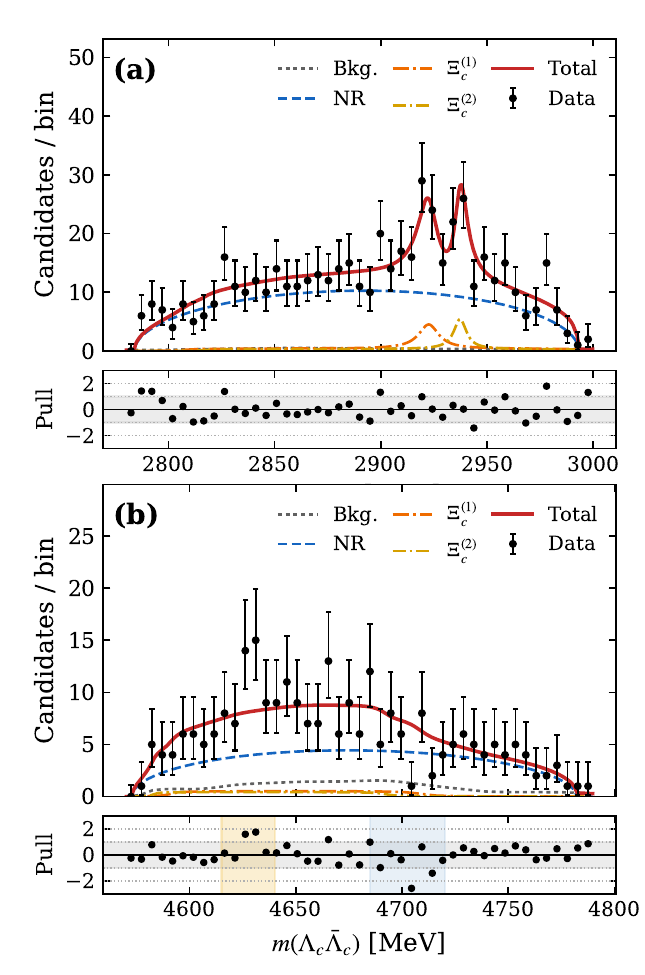}
\caption{Coherent baseline description of the two invariant-mass projections: (a) the $m(\Lambda_c^+K_S^0)$ spectrum and (b) the $m(\Lambda_c^+\bar{\Lambda}_c^-)$ spectrum. The red curves show the complete model, while the blue, orange, yellow, and gray curves denote the nonresonant, $\Xi_c^{(1)}$, $\Xi_c^{(2)}$, and fixed LHCb background contributions, respectively. The lower panels show the pulls, $p_i=(n_i-\mu_i)/\sigma_i$, evaluated using the corresponding side of the asymmetric uncertainty as defined in Eq.~\eqref{eq:fit-projection-chi2}; the grey band marks $\pm1\sigma$ and the shaded intervals in panel (b) indicate the 4.63~GeV and 4.7~GeV regions of interest.}
\label{fig:background-comparison}
\end{figure}

The coherent baseline in the two projections is shown in Fig.~\ref{fig:background-comparison}.  Its fitted amplitude parameters are $A_{\rm NR}=12.20$ and $(C_{\Xi_c^{(1)}},C_{\Xi_c^{(2)}})=(2.03-10.97i,-1.67-12.75i)$. We obtain {$(\chi_1^2,\chi_2^2)$}=$(25.35,24.90)$ for the coherent baseline.

The pull in each bin is defined as $p_i=(n_i-\mu_i)/\sigma_i$, with $\sigma_i$ chosen from the upper or lower quoted uncertainty on the side of the fitted value, as in Eq.~\eqref{eq:fit-projection-chi2}.  In the pull panels of Fig.~\ref{fig:background-comparison}, the grey band marks the $\pm1\sigma$ range: for a model that describes the data well, the pulls are expected to scatter within this band, whereas persistent excursions beyond it indicate localized deviations from the baseline description.  In panel (b), the shaded vertical intervals highlight the 4.63~GeV and 4.7~GeV regions: near 4.63~GeV several bins of the $m_{12}$ spectrum lie above the baseline by almost $2\sigma$, while near 4.7~GeV a single bin dips below the baseline by more than $2\sigma$.  These deviations are precisely the regions in which additional $\Lambda_c^+\bar{\Lambda}_c^-$-channel amplitudes will be introduced below.

In the charmonium family, the pseudoscalar $\eta_c$ state provides the lowest partial-wave configuration for this process. Both $B\to K\eta_c$ and $\eta_c\to\Lambda_c^+\bar{\Lambda}_c^-$ can proceed in an $S$ wave, whereas $B\to K\psi$ requires a $P$ wave.\footnote{Here, $\eta_c$ and $\psi$ denote pseudoscalar and vector charmonia, respectively.} Higher-spin states involve even higher partial waves in their production. We therefore examine whether the masses and widths of excited $\eta_c$ states can account for the two fitted enhancement regions.

\subsection{{Masses of $\eta_c(6S)$ and $\eta_c(7S)$ in the modified GI model}}
\label{sec:theory}

We calculate the charmonium spectrum in the modified Godfrey--Isgur (MGI) model, in which the linear confining interaction of the original GI model~\cite{Godfrey:1985xj} is replaced by the screened form adopted in Ref.~\cite{Wang:2020yaa}.  Apart from this replacement and the parameter sets specified below, the spin-dependent interactions, smearing prescription, momentum-dependent corrections, and remaining model parameters are kept the same as in our previous studies of highly excited charmonia~\cite{Wang:2019,Wang:2020yaa,Liu:2024prd}.  We therefore summarize only the ingredients needed for the present calculation.

Experience from our previous spectroscopic studies indicates that screening effects become increasingly important for highly excited charmonia, particularly for levels with $n\geq5$.  To account phenomenologically for string breaking at large interquark separations, we therefore follow Refs.~\cite{Wang:2019,Wang:2020yaa,Liu:2024prd} and replace the linear term $br$ by

\begin{equation}
    b r \;\longrightarrow\; \frac{b}{\mu}\bigl(1 - e^{-\mu r}\bigr),
    \label{eq:screen_replace}
\end{equation}
The full screened confining potential before smearing can be written as $V_{\rm conf}(r)=c+(b/\mu)(1-e^{-\mu r})$. Here, $b$ is the coefficient of the linear confining potential and $c$ is its additive constant, with units of GeV$^2$ and GeV, respectively. The Eq.~\eqref{eq:screen_replace} approaches $b/\mu$ as $r\to\infty$ and thereby mimics the saturation of the confining interaction due to string breaking.

The sensitivity to the screening parameter increases with radial excitation. As shown in Ref.~\cite{Wang:2020yaa}, varying $\mu$ from 0.11 to 0.15~GeV changes the predicted $\psi(6S)$ and $\psi(7S)$ masses by about 98 and 144~MeV, respectively, while the established lower states are less  sensitive.  In view of the sizeable model uncertainty in this highly excited region, we do not base the spectroscopic fit on a single value of $\mu$.  We take $\mu=0.12$~GeV and $(b,c)=(0.238~\mathrm{GeV}^2,-0.337~\mathrm{GeV})$ as the central parameter set and use $\mu=0.11$ and 0.13~GeV to define the variation range, while keeping the remaining model parameters identical to those in Ref.~\cite{Wang:2020yaa}.

For the two endpoint sets, $b$ and $c$ are refitted to the experimental masses of the established charmonium states, following the procedure of Ref.~\cite{Wang:2020yaa}.  This procedure gives $(b,c)=(0.230~\mathrm{GeV}^2,-0.329~\mathrm{GeV})$, $(0.238~\mathrm{GeV}^2,-0.337~\mathrm{GeV})$, and $(0.246~\mathrm{GeV}^2,-0.344~\mathrm{GeV})$ for $\mu=0.11$, 0.12, and 0.13~GeV, respectively. These parameter sets reproduce the experimental masses of the established charmonium states.  The variation over $\mu=0.12\pm0.01$~GeV is propagated through the predicted masses and initial-state wave functions.  The resulting mass and width ranges are then used to select the representative inputs for the line-shape fits.\par

The spectra and strong-decay properties of other high-lying charmonia relevant to this mass region have been calculated in our previous     studies~\cite{Wang:2019,Wang:2020yaa,Liu:2024prd,Peng:2024}. Here Table~\ref{tab:gi} lists only the pseudoscalar states used to compare the fitted regions with the screened spectrum in the present study.

Following the GI numerical prescription~\cite{Godfrey:1985xj}, we solve the Hamiltonian eigenvalue problem with the screened potential by expanding the radial wave function in a finite harmonic-oscillator basis.  The basis is chosen sufficiently large that the relevant energy eigenvalues are stable under further enlargement.  The resulting spatial wave functions are then used as input to the QPC decay calculation in Sec.~\ref{sec:qpc-widths}.

The masses of the low-lying states change only mildly over the three parameter sets, whereas the higher radial excitations show a larger model systematic variation.  Table~\ref{tab:gi} presents the calculated spectrum used to examine whether the structures near 4.63 and 4.7~GeV may be associated with $\eta_c(6S)$ and $\eta_c(7S)$, respectively.

\begin{table}[htbp]
\caption{{Masses of $\eta_c (nS)$ states predicted by the screened GI model, in MeV, for the three parameter sets associated with $\mu=0.11$, 0.12, and 0.13~GeV. Experimental masses are taken from Ref.~\cite{ParticleDataGroup:2026aaa}. A dash indicates that no experimentally established mass is available.}}
\label{tab:gi}
\centering
{
\setlength{\tabcolsep}{2.5pt}
\renewcommand{\arraystretch}{1.25}
\begin{tabular}{lcccc}
\toprule
$\eta_c(nS)$ & $\mu=0.11$ GeV & $\mu=0.12$ GeV & $\mu=0.13$ GeV &
{Experiment} \\ 
\midrule
$\eta_c(1S)$ & 2984.3 & 2983.9 & 2984.0 & {$2984.1 \pm 0.3$} \\
$\eta_c(2S)$ & 3626.4 & 3628.9 & 3631.5 & {$3637.8 \pm 0.6$} \\
$\eta_c(3S)$ & 4002.3 & 4002.2 & 4001.9 & {--} \\
$\eta_c(4S)$ & 4269.4 & 4263.4 & 4256.9 & {--} \\
$\eta_c(5S)$ & 4472.5 & 4458.3 & 4443.3 & {--} \\
$\eta_c(6S)$ & 4630.7 & 4606.6 & 4581.4 & {--} \\
$\eta_c(7S)$ & 4755.1 & 4719.6 & 4682.7 & {--} \\
\bottomrule
\end{tabular}
}
\end{table}

Varying the MGI parameter set gives an $\eta_c(6S)$ mass range of 4581.4$\sim$4630.7~MeV.  The event accumulation near 4.63~GeV lies at the upper end of this range and is closest to the prediction obtained with $\mu=0.11$~GeV; the central set with $\mu=0.12$~GeV instead gives 4606.6~MeV.  The corresponding $\eta_c(7S)$ range is $4682.7\sim 4755.1$~MeV, and the central prediction of 4719.6~MeV lies in the region of the possible higher-mass structure around 4.7~GeV. We next examine the predicted open-charm widths to compare the decay properties of the two states with the fitted structures.

\subsection{{Open-charm decay widths of $\eta_c(6S)$ and $\eta_c(7S)$}}
\label{sec:qpc-widths}

We study the open-charm decays of $\eta_c(6S)$ and $\eta_c(7S)$ in the quark-pair-creation (QPC) model, also known as the $^3P_0$ model~\cite{Micu:1968mk,LeYaouanc:1972vsx,Ackleh:1996yt,Barnes:2005pb}. In this framework, the decay $A\to B+C$ proceeds through the creation of a light $q\bar q$ pair with the vacuum quantum numbers $J^{PC}=0^{++}$.  We use the standard QPC transition operator to calculate the helicity amplitudes and transform them into partial-wave amplitudes with the Jacob--Wick formula ~\cite{Jacob:1959at,LeYaouanc:1972vsx,Blundell:1996}, following the same convention as in our previous studies~\cite{Wang:2020yaa,Liu:2024prd}.  With the amplitude normalization adopted in the numerical calculation, the partial width is written as
\begin{equation}
    \Gamma_{A\to BC} = \frac{\pi}{4}\frac{|\mathbf{P}|}{M_A^2}\,  \sum_{J,L}|\mathcal{M}^{JL}(\mathbf{P})|^2,
    \label{eq:QPC_width}
\end{equation}
where $\mathbf{P}$ denotes the three-momentum of either final-state meson in the rest frame of $A$, and $J$ and $L$ are the total and orbital angular momenta of the $BC$ system, respectively.  The explicit QPC transition operator and helicity-amplitude expressions are given in Ref.~\cite{Blundell:1996}.  Following Ref.~\cite{Wang:2020yaa}, we take the light-quark pair-creation strength to be $\gamma_{u,d}=5.84$, as used in the QPC analysis of highly excited vector charmonia around 4.6~GeV.  The creation of an $s\bar s$ pair is suppressed by setting $\gamma_s=\gamma_{u,d}/\sqrt{3}$.  

The masses and spatial wave functions of the final charmed and charmed-strange mesons are taken from the model calculations of Refs.~\cite{Song:2015D,Song:2015Ds}.  For the ground states we use $m_D=1870$, $m_{D^*}=2007$, $m_{D_s}=1970$, and $m_{D_s^*}=2112$~MeV.  The masses and wave functions of the excited $D_J^{(*)}$ and $D_{sJ}^{(*)}$ states are taken from the same references.

The dominant OZI-allowed two-body open-charm decay modes of $\eta_c(6S)$ are collected in Table~\ref{tab:qpc}.  For the central parameter set, the sum of the included two-body open-charm widths is $\Gamma_{\rm open}[\eta_c(6S)]\approx27.90$~MeV.  The largest contributions arise from $D\bar D_0^*(2300)$, $D\bar D_2^*(2460)$, and $D^*\bar D^*$, with partial
widths of 12.60, 5.40, and 4.70~MeV, respectively. For each $(\mu,b,c)$ parameter set, the corresponding MGI mass and radial wave function of the initial $\eta_c$ state are used in the QPC calculation. Changing the parameter set therefore modifies not only the available phase space through the initial-state mass, but also the decay amplitude through the wave-function overlap.

Repeating the full calculation gives $\Gamma_{\rm open}=33.80$~MeV for the $\mu=0.11$~GeV set and $\Gamma_{\rm open}=24.20$~MeV for the $\mu=0.13$~GeV set.  Together with the
central result at $\mu=0.12$~GeV, these values give a {model-induced uncertainty} of $24.20\sim 33.80$~MeV.  Despite the number of kinematically allowed open-charm channels, the summed width remains at the level of several tens of MeV, which can be attributed in part to nodal cancellations in the wave-function overlaps. For $\eta_c(6S)$, channels with widths below 1~MeV are not listed individually, except for $D\bar D^*$, whose strongly suppressed width of about 1.10~keV is retained for illustration.

\begin{table*}[htbp]
\centering
\setlength{\tabcolsep}{3pt}
\renewcommand{\arraystretch}{1.2}
\caption{{Dominant open-charm decay modes of $\eta_c(6S)$ and $\eta_c(7S)$.  The central widths are calculated with $\gamma=5.84$ and the initial-state MGI wave functions for the $\mu=0.12$~GeV parameter set.  The superscript (subscript) gives the signed shift obtained with the $\mu=0.11$ ($0.13$) parameter set and represents the screening-induced model systematic variation.  Unlisted subleading modes are grouped under ``Other channels,'' whereas $D\bar D^*$ is shown separately to illustrate its node sensitivity. The fractions are evaluated relative to the central summed open-charm
width.}}
\label{tab:qpc}
\begin{tabular}{lcc@{\hspace{1.2em}}lcc}
\toprule
\multicolumn{3}{c}{$\eta_c(6S)$} &
\multicolumn{3}{c}{$\eta_c(7S)$}\\
\cmidrule(r){1-3}\cmidrule(l){4-6}
Decay channel & $\Gamma$ [MeV] & Fraction &
Decay channel & $\Gamma$ [MeV] & Fraction\\
\midrule
$D\bar D_0^*(2300)$ & $12.60^{+0.50}_{-0.90}$ & 45\% &
$D\bar D_0^*(2300)$ & $9.87^{+0.07}_{-0.58}$ & 31\%\\
$D\bar D_2^*(2460)$ & $5.40^{+0.40}_{-1.20}$ & 19\% &
$D^*\bar D_1(2420)$ & $5.98^{+3.82}_{-3.41}$ & 19\%\\
$D^*\bar D^*$ & $4.70^{-1.10}_{+1.00}$ & 17\% &
$D^*\bar D_1(2430)$ & $4.20^{+1.93}_{-2.41}$ & 13\%\\
$D^*\bar D_1(2420)$ & $1.60^{+2.00}_{-1.10}$ & 6\% &
$D\bar D_2^*(2460)$ & $3.55^{-0.65}_{-0.19}$ & 11\%\\
$D^*\bar D_1(2430)$ & $1.70^{+2.20}_{-1.30}$ & 6\% &
$D\bar D^*(2600)$ & $2.69^{+1.34}_{-1.77}$ & 8\%\\
$D_s\bar D_{s0}^*(2317)$ & $1.10^{+0.20}_{-0.10}$ & 4\% &
$D^*\bar D^*$ & $2.50^{-0.95}_{+1.12}$ & 8\%\\
$D\bar D^*$ & $0.0011^{+0.20}_{+0.15}$ & $<0.01\%$ &
$D^*\bar D_2^*(2460)$ & $1.45^{+2.19}_{-1.33}$ & 4\%\\
{Other channels} & {$0.80^{+1.50}_{-0.25}$} & {3\%} &
$D_s\bar D_{s0}^*(2317)$ & $0.99^{+0.13}_{-0.18}$ & 3\%\\
$\cdots$ & $\cdots$&$\cdots$ & $D\bar D^*$ & $0.0810^{+0.48}_{-0.01}$ & 0.25\%\\
$\cdots$ & $\cdots$& $\cdots$& Other channels & $1.00^{+1.32}_{-0.70}$ & 3\%\\
\midrule
{Open-charm sum} & $27.90^{+5.90}_{-3.70}$ & 100\% &
{Open-charm sum} & $32.31^{+9.67}_{-9.47}$ & 100\%\\
\bottomrule
\end{tabular}
\end{table*}

For $\eta_c(7S)$, the dominant central partial widths are associated with $D\bar D_0^*(2300)$, $D^*\bar D_1(2420)$, $D^*\bar D_1(2430)$, and $D\bar D_2^*(2460)$, for which the predicted widths are 9.87, 5.98, 4.20, and 3.55~MeV, respectively.  The summed open-charm width is 32.31~MeV for the central parameter set and spans $22.84\sim 41.98$~MeV over the three sets, which we use as its model-induced uncertainty.  The $D\bar D^*$ mode remains below 1~MeV throughout this variation, with widths of 0.557, 0.081, and 0.070~MeV for the $\mu=0.11$, 0.12, and 0.13~GeV sets, respectively.

For direct comparison with the line-shape fits in the next subsection, the three parameter sets give $M[\eta_c(6S)]=4581.4\sim 4630.7$~MeV and $\Gamma[\eta_c(6S)]=24.20\sim 33.80$~MeV.  For $\eta_c(7S)$, the corresponding ranges are $4682.7\sim 4755.1$~MeV and $22.84\sim 41.98$~MeV.  The summed open-charm widths are used as proxies for the resonance total widths, under the assumption that decay modes outside the two-body open-charm calculation give subleading contributions.

\subsection{Spectroscopic fit to the $\Lambda_c^+\bar{\Lambda}_c^-$ and $\Lambda_c^+K_S^0$ invariant-mass spectra}
\label{sec:data-fit-results}

\begingroup

Building on the baseline fit of Sec.~\ref{sec:coherent-model}, we now extend the amplitude with charmonium resonances and fit the two spectra with {the $\eta_c(6S)$ and $\eta_c(7S)$ parameters fixed to representative values selected from the spectroscopic ranges above}.  With the coherent baseline adopted as the reference, we compare three amplitude models:
\begin{equation}
\begin{aligned}
 H_0 &: {\rm NR}+\Xi_c,\\
 H_6 &: {\rm NR}+\Xi_c+\eta_c(6S),\\
 H_{67} &: {\rm NR}+\Xi_c+\eta_c(6S)+\eta_c(7S),
\end{aligned}
\label{eq:fit-hypotheses}
\end{equation}
where $H_6$ denotes the model with $\eta_c(6S)$ and $H_{67}$ the model with both $\eta_c(6S)$ and $\eta_c(7S)$. We determine $C_{\Xi_c^{(1)}}$ and $C_{\Xi_c^{(2)}}$ from the coherent baseline.  In the $\eta_c(6S)$ and $\eta_c(6S)+\eta_c(7S)$ fits, their relative magnitude and phase are kept fixed, while a common real factor $s_{\Xi}$ is allowed to vary; we set $C_{\Xi_j}\to s_{\Xi}C_{\Xi_j}$.  This keeps the two $\Xi_c$ line shapes consistent with the baseline fit while limiting the number of additional parameters. The $\eta_c(6S)$ term describes the enhancement near 4.63~GeV, and $\eta_c(7S)$ probes the higher-mass region.  The complete
amplitude is
\begin{equation}
 \begin{aligned}
 \mathcal M(s_{12},s_{13}) &=\mathcal M_{\Xi_c}(s_{13}) +\sum_{n}C_{\eta_c(nS)}{\rm RBW}_{\eta_c(nS)}(s_{12}),\\
 \mathcal I_{\rm sig}^{\rm coh} &=|\mathcal M(s_{12},s_{13})|^2,
 \end{aligned}
 \label{eq:fit-coherent-x-intensity}
\end{equation}
where the sum runs over the included $\eta_c(nS)$ states, $n=6$ or $7$. Thus the $\eta_c(nS)$ terms interfere with the nonresonant and both $\Xi_c$ amplitudes, as well as with each other when both are present.

Unconstrained scans also produce minima with widths of only a few MeV. These solutions generate sharp peak--dip variations controlled by a few adjacent 5~MeV bins, and such small widths are atypical of highly excited charmonia.  The subsequent spectroscopic analysis therefore adopts solutions with widths on the tens-of-MeV scale.

For the spectroscopic tests, the resonance masses and widths are fixed to representative values selected from the ranges obtained by varying $\mu=0.12\pm0.01$~GeV: $(M_1,\Gamma_1)=(4629.00,24.20)$~MeV for $\eta_c(6S)$ and $(M_2,\Gamma_2)=(4718.84,22.84)$~MeV for $\eta_c(7S)$. The masses follow the line-shape locations identified in the preliminary scans, while the widths are taken from the lower edges of the corresponding QPC ranges.  These four values are specified before the amplitude coefficients are optimized.

The two projections are constructed from the same event sample.  The experiment does not provide the statistical relation between the two projections, so we report the two contributions separately and use their sum to compare the line shapes. For each model, $A_{\rm NR}$, $s_{\Xi}$, and the complex resonance coefficients are reoptimized by minimizing {$\chi_1^2+\chi_2^2$}; {$\nu_1$} is updated by the yield-matching prescription above. The two contributions $\chi_1^2$ and $\chi_2^2$ are quoted separately throughout.

The relativistic Breit--Wigner form used in the fit requires an energy-dependent width $\Gamma(s)$, whereas the QPC calculation provides only the total open-charm width evaluated at the resonance mass, $\Gamma(M^2)$. We therefore use the fixed QPC width as the on-shell input and model its energy dependence phenomenologically with the $S$-wave $\Lambda_c\bar\Lambda_c$ momentum in Eq.~\eqref{eq:fit-running-width}. As a robustness check, replacing this running width by a constant width changes the total $\chi^2$ by only $0.09$ for $H_6$ and $0.22$ for $H_{67}$, indicating that the fits are insensitive to this prescription.

As a robustness check, we also allow the two complex $\Xi_c$ coefficients to vary independently in the corrected fit including both $\Lambda_c K_S^0$ combinations.  This gives $(\chi_1^2,\chi_2^2)$=$(25.89,20.42)$ for $H_6$ and $(23.17,16.74)$ for $H_{67}$, corresponding to reductions of only $0.05$ and $0.86$, respectively.  The optimized $\eta_c$ coefficients remain on the same overall scale, and the additional $\Xi_c$ freedom does not materially change the line-shape description.

Adding the fixed $\eta_c(6S)$ amplitude changes $(\chi_1^2,\chi_2^2)$ from $(25.35,24.90)$ for the coherent baseline to $(25.99,20.37)$.  The improvement is therefore concentrated in the $m_{12}$ projection containing the 4.63~GeV enhancement.  Including the fixed $\eta_c(7S)$ amplitude gives {$(24.27,16.51)$} and improves the description of the higher-mass part of the same projection.  The small change in $\chi_1^2$ shows that both additions preserve the description of the $\Lambda_c^+K_S^0$ spectrum.  Table~\ref{tab:data-driven-hypotheses} collects the coherent baseline and the two spectroscopy-guided fixed-parameter fits, together with their optimized amplitude parameters.

The two projections for the $\eta_c(6S)$ fit are shown in Fig.~\ref{fig:structure-solutions}(a,b).  The component curves show the diagonal intensities, while their interference is included in the red total curve.  With the 4.63~GeV region described, we next test the higher-$m_{12}$ region by adding $\eta_c(7S)$.  The two projections for the combined solution are shown in Fig.~\ref{fig:structure-solutions}(c,d).

\begin{table*}[htbp]
\centering
\small
\setlength{\tabcolsep}{3pt}
\caption{Baseline and spectroscopy-guided fixed-parameter fit results. Masses and widths are in MeV.  The first row is the coherent baseline; the second and third rows include the fixed $\eta_c(6S)$ and $\eta_c(6S)+\eta_c(7S)$ amplitudes, respectively.  $A_{\rm NR}$ is the nonresonant amplitude, $s_\Xi$ rescales the baseline $\Xi_c$ amplitudes, and $C_1$ ($C_2$) is the complex coefficient of the $\eta_c(6S)$ [$\eta_c(7S)$] amplitude in Eq.~\eqref{eq:fit-coherent-x-intensity}. With the Breit--Wigner normalization used above, $A_{\rm NR}$, $s_\Xi$, $C_1$, and $C_2$ are dimensionless. {The $\chi_1^2$ and $\chi_2^2$ values are the sums defined in Eq.~\eqref{eq:fit-projection-chi2}, without division by the number of degrees of freedom.}}
\label{tab:data-driven-hypotheses}
\begin{tabular}{lrrrrrrrrrrrr}
\toprule
Fit & $M_1$ [MeV] & $\Gamma_1$ [MeV] &
$M_2$ [MeV] & $\Gamma_2$ [MeV] &
$A_{\rm NR}$ & $s_\Xi$ & ${\rm Re}\,C_1$ & ${\rm Im}\,C_1$ &
${\rm Re}\,C_2$ & ${\rm Im}\,C_2$ & {$\chi_1^2$} & {$\chi_2^2$}\\
\midrule
$H_0$ coherent & -- & -- & -- & -- & {12.20} & 1.00 & -- & -- & -- & -- & {25.35} & {24.90}\\
\textbf{$\boldsymbol{H}_6$} &
\textbf{4629.00} & \textbf{24.20} & \textbf{--} & \textbf{--} &
{\textbf{11.82}} & {\textbf{0.84}} &
{\textbf{$-1.34$}} & {\textbf{$-4.55$}} & -- & -- &
{\textbf{25.99}} & {\textbf{20.37}}\\
\textbf{$\boldsymbol{H}_{67}$} &
\textbf{4629.00} & \textbf{24.20} &
\textbf{4718.84} & \textbf{22.84} &
{\textbf{12.22}} & {\textbf{0.96}} &
{\textbf{$-3.00$}} & {\textbf{$-4.48$}} &
{\textbf{$-4.46$}} & {\textbf{4.14}} &
{\textbf{24.27}} & {\textbf{16.51}}\\
\bottomrule
\end{tabular}
\end{table*}

\begin{figure*}[t]
\centering
\includegraphics[width=0.98\textwidth]{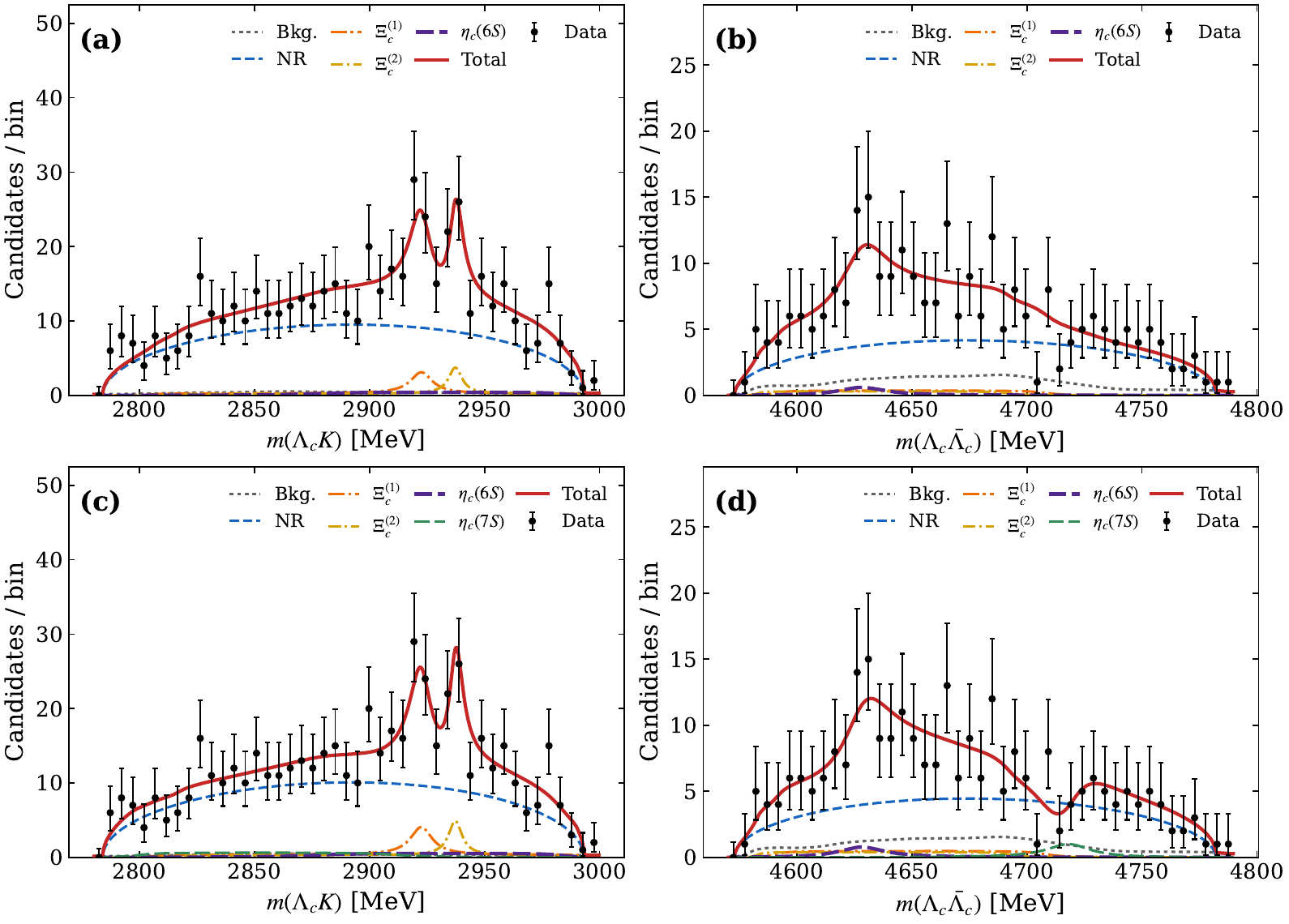}
\caption{Spectroscopic fits in the two projections.  Panels (a) and (b) show the $\eta_c(6S)$ solution, $(M_1,\Gamma_1)=(4629.00,24.20)$~MeV, in the $m(\Lambda_c^+K_S^0)$ and $m(\Lambda_c^+\bar\Lambda_c^-)$ spectra, respectively.  Panels (c) and (d) show the solution with both $\eta_c(6S)$ and $\eta_c(7S)$, $(M_1,\Gamma_1;M_2,\Gamma_2)=(4629.00,24.20;4718.84,22.84)$~MeV, in the same two projections.  Red curves are the complete models; the other curves show the diagonal contributions.  Here $\Xi_c^{(1)}$ and $\Xi_c^{(2)}$ denote the two established $\Xi_c$ contributions, and ``Bkg.'' is the fixed LHCb background.}
\label{fig:structure-solutions}
\end{figure*}

In the fixed $H_{67}$ solution, the relatively small direct $\eta_c(7S)$ contribution is accompanied by a larger destructive interference term, producing a localized dip around 4.71~GeV. This shows that the spectroscopic results remain compatible with the fit results. Taken together, the mass and width comparisons make the $\eta_c(6S)$ a plausible conventional candidate for the structure near 4.63~GeV, while the $\eta_c(7S)$ contribution provides an exploratory test of the higher-mass line shape. More precise measurements will be required to determine its line shape and physical origin.

\endgroup

\section{Baryonic decays and weak production of high-lying $\eta_c$ states}
\label{sec:phenomenology}

The preceding analysis provides the mass and width inputs used below for the two
high-lying $\eta_c$ states considered above.  We now follow these inputs through the complete decay chains

\begin{equation*}
	B^0\to K_S^0\eta_c(nS)\to K_S^0\Lambda_c^+\bar{\Lambda}_c^-, \qquad n=6,7.
\end{equation*}
We first evaluate the baryonic transitions $\eta_c(nS)\to\Lambda_c^+\bar{\Lambda}_c^-$ within a common hadron-loop mechanism and then estimate $B^0\to K_S^0\eta_c(nS)$ using naive factorization.  Finally, we use the fitted $\eta_c(6S)$ fraction to estimate the branching fraction of the complete $6S$ decay chain and compare it with theoretical estimates.

\subsection{Baryonic decays of $\eta_c(6S)$ and $\eta_c(7S)$}
\label{sec:branching}

The QPC calculation in Sec.~\ref{sec:qpc-widths} describes open-charm meson-pair channels through the creation of a single light $q\bar q$ pair. Direct production of  $\Lambda_c^+\bar{\Lambda}_c^-$ from the initial $c\bar c$ state requires two light $q\bar q$ pairs and therefore lies outside this one-pair QPC treatment.  We estimate the baryonic channel through a hadronic-loop mechanism.

\begin{figure}[t]
\centering
\includegraphics[width=\columnwidth]{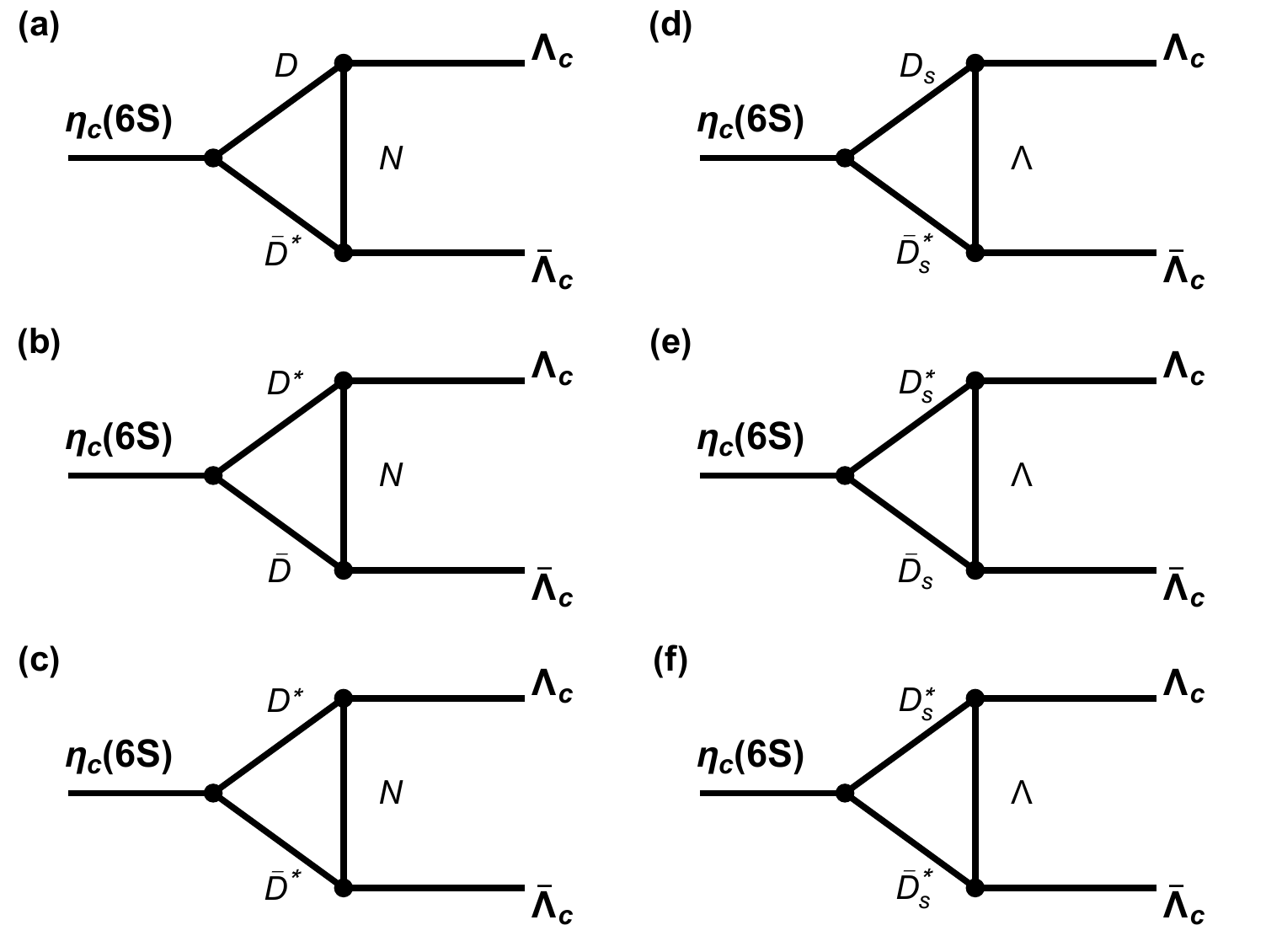}
\caption{Hadronic-loop diagrams included in the calculation of $\eta_c(6S)\to\Lambda_c^+\bar{\Lambda}_c^-$.  The same loop topologies are used for $\eta_c(7S)$.  The left column shows nonstrange open-charm loops, $D^{(*)}\bar D^{(*)}$, with an exchanged nucleon $N$, whereas the right column shows strange open-charm loops, $D_s^{(*)}\bar D_s^{(*)}$, with an exchanged hyperon ($\Lambda$).  Charge-conjugate processes are included implicitly.}
\label{fig:loop_LcLc}
\end{figure}

At leading twist in perturbative QCD, the decays $\eta_c(nS)\to\Lambda_c^+\bar{\Lambda}_c^-$ are suppressed by the hadron-helicity selection rule~\cite{Brodsky:1981}.  Nevertheless, long-distance rescattering can contribute.  We describe this contribution with the hadron-loop mechanism~\cite{Liu:2010}, in which intermediate open-charm meson pairs $M_1M_2$ rescatter into the baryon-antibaryon final state. The triangle diagram gives the amplitude
\begin{multline}
\mathcal{A}_{\rm loop}(P^2) =\sum_{M_1M_2} \int \frac{d^4q}{(2\pi)^4} \frac{V_{M_1M_2}\,T_{M_1B}\,T_{M_2B}} {(q_1^2-m_1^2)(q_2^2-m_2^2)} \frac{\mathcal{F}^2(q^2)}{q^2-m_B^2},
\label{eq:hadron_loop}
\end{multline}
where the product $T_{M_1B}T_{M_2B}$ denotes the complete spin-dependent meson-baryon numerator.  It includes the two effective vertices of Eq.~\eqref{eq:L_Dstar}, the external baryon spinors, the numerator of the exchanged-baryon propagator, and, for an intermediate vector meson, the corresponding spin-1 propagator numerator.  The explicit numerator structures are given in Appendix~\ref{app:etac}.  Here, $B$ denotes the exchanged ground-state baryon ($N$ or $\Lambda$), while $q_1$, $q_2$, and $q$ are the momenta of the two intermediate mesons and the exchanged baryon, respectively.  Figure~\ref{fig:loop_LcLc} shows the Feynman diagrams contributing to the hadron-loop amplitude.

At $P^2=m_{\eta_c(nS)}^2$, the partial width and branching fraction are evaluated as
\begin{equation}
 \Gamma_{\rm loop}= \frac{|\boldsymbol p_{\Lambda_c}|}{8\pi m_{\eta_c(nS)}^2} \sum_{\rm spins}|\mathcal A_{\rm loop}|^2, \qquad
 \mathcal B_{\rm loop}=\frac{\Gamma_{\rm loop}}{\Gamma_{\eta_c(nS)}},
 \label{eq:loop-width}
\end{equation}
where $\boldsymbol p_{\Lambda_c}$ is the three-momentum of either final-state baryon in the $\eta_c(nS)$ rest frame. For the pseudoscalar $\eta_c$, the couplings to open-charm meson pairs take the forms ~\cite{Casalbuoni:1996pg}
\begin{equation}
\mathcal{L}_{\eta_c D D^*}
= 2i g_{\eta_c D D^*}\,\eta_c \bigl(D^{*\mu}\partial_\mu\bar D-\bar D^{*\mu}\partial_\mu D\bigr),
\label{eq:L_DDstar}
\end{equation}
\begin{equation}
\mathcal{L}_{\eta_c D^* D^*} = 2 g_{\eta_c D^* D^*}\, \partial^\lambda\eta_c\, \bigl(\bar D^{*\nu}\partial^\mu D^{*\alpha}- D^{*\alpha}\partial^\mu\bar D^{*\nu}\bigr)\, \varepsilon_{\alpha\nu\mu\lambda},
\label{eq:L_DstarDstar}
\end{equation}
where $D$ and $\bar D$ denote the pseudoscalar meson and antimeson, while $D^*$ and $\bar D^*$ denote their vector partners.  The same forms are used for the $D_s^{(*)}$ channels.  We determine the couplings by matching the partial widths obtained from these effective vertices to the QPC results in Sec.~\ref{sec:qpc-widths}.  For $\eta_c(6S)$, the resulting couplings are  $g_{\eta_c D D^*}=0.0026$, $g_{\eta_c D^*D^*}=0.036$~GeV$^{-1}$, $g_{\eta_c D_sD_s^*}=0.0485$, and  $g_{\eta_c D_s^*D_s^*}=0.0074$~GeV$^{-1}$.  The corresponding $7S$ values are $g_{\eta_c D D^*}=0.020$, $g_{\eta_c D^*D^*}=0.023$~GeV$^{-1}$, $g_{\eta_c D_sD_s^*}=0.031$, and $g_{\eta_c D_s^*D_s^*}=0.0047$~GeV$^{-1}$.  The relative signs of these couplings are fixed by the heavy-quark-symmetry convention of the effective Lagrangian, and their common overall sign is chosen to be positive.  For the loop kinematics, we use the masses in the fixed $H_{67}$ fit, $m_{\eta_c(6S)}=4.629$~GeV and $m_{\eta_c(7S)}=4.719$~GeV. We use rounded total widths of 28 and 32~MeV to convert the $6S$ and $7S$ partial widths into branching fractions.

The rescattering kernel contains the $t$-channel exchange of ground-state octet baryons.  For nonstrange intermediate states ($D^{(*)}\bar D^{(*)}$), we retain nucleon exchange, described by the effective Lagrangian
\begin{align}
\mathcal{L}_{D^{(*)}\Lambda_c N} =& i g_{D\Lambda_c N}\,\bar N\gamma_5\Lambda_c D + g_{D^*\Lambda_c N}\,\bar N\gamma_\mu\Lambda_c D^{*\mu} \nonumber \\ 
&+ \frac{\kappa_{D^*\Lambda_c N}}{m_{\Lambda_c}+m_N}\, \bar N\sigma_{\mu\nu}\Lambda_c\,\partial^\nu D^{*\mu} +{\mathrm{H.c.}},
\label{eq:L_Dstar}
\end{align}
where $g_{D\Lambda_c N}$, $g_{D^*\Lambda_c N}$, and $\kappa_{D^*\Lambda_c N}$ denote the pseudoscalar, vector, and tensor coupling constants, respectively, with values taken from Ref.~\cite{Qian:2021}: $g_{D\Lambda_c N}=13.8$, $g_{D^*\Lambda_c N}=-7.9$, and $\kappa_{D^*\Lambda_c N}=4.7$.  For strange intermediate states ($D_s^{(*)}\bar D_s^{(*)}$), we retain $\Lambda$ exchange. The interaction has the analogous Lorentz structure, and the coupling constants are obtained from the SU(3)-flavor relation: $g (\kappa)_{D^{(*)}\Lambda_cN}=-\sqrt{3/2}\,g (\kappa)_{D_s^{(*)}\Lambda_c\Lambda}$. The masses entering the loop integrals are $m_D=1.870$~GeV, $m_{D^*}=2.010$~GeV, $m_{D_s}=1.968$~GeV, $m_{D_s^*}=2.112$~GeV, $m_N=0.939$~GeV, and $m_\Lambda=1.116$~GeV~\cite{ParticleDataGroup:2026aaa}.

We introduce an effective dipole form factor
\begin{equation}
    \mathcal{F}(q^2;\Lambda) = \left(\frac{\Lambda^2-m^2}{\Lambda^2-q^2}\right)^2, \qquad \Lambda = m + \alpha\Lambda_{\rm QCD},
    \label{eq:ff}
\end{equation}
to regulate the exchanged baryon line, where $m$ is the mass of the ground-state baryon ($N$ or $\Lambda$) and $\Lambda_{\rm QCD}=0.22$~GeV. For the charmed baryon pair decay of the $6S$-wave-dominated mixed-vector charmonium considered in Ref.~\cite{Qian:2021}, the inferred $\Lambda_c\bar\Lambda_c$ rate is reproduced with $\alpha=1.9\sim 2.3$ under the assumed dilepton width, whereas the other mixed-vector charmonium state requires larger values of $\alpha$.  By transferring these constraints to the pseudoscalar partner states considered here, we use $\alpha=1.0\sim 3.0$ as a phenomenological scan.

\begin{figure}[t]
\centering
\includegraphics[width=\columnwidth]{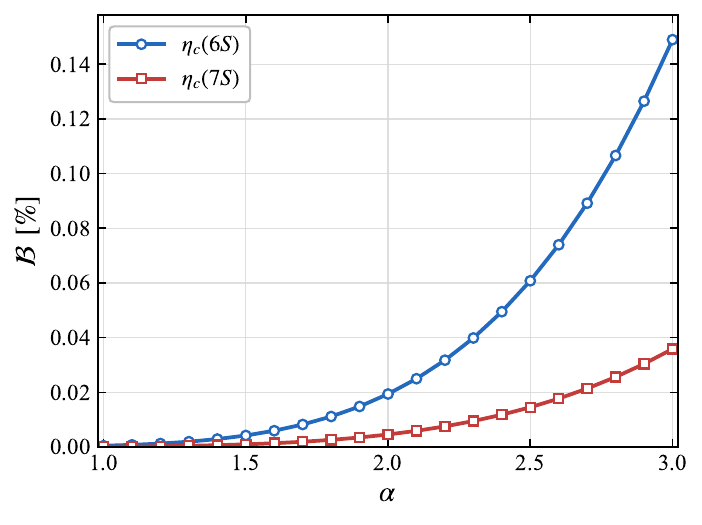}
\caption{{Hadron-loop branching fractions for $\eta_c(6S,7S)\to\Lambda_c^+\bar{\Lambda}_c^-$ as functions of $\alpha$.}}
\label{fig:LcLc}
\end{figure}

Figure~\ref{fig:LcLc} shows the cutoff dependence.  The $6S$ branching fraction reaches $10^{-4}$ at $\alpha\simeq1.8$, whereas the smaller $7S$ result reaches the same scale at $\alpha\simeq2.4$.  Thus, both branching fractions are of order $10^{-4}$ over the upper part of the commonly used cutoff range.  Their absolute widths or branching ratios change rapidly with $\alpha$, whereas their ratio is considerably more stable:
\begin{equation}
	\frac{\mathcal{B}^{\rm loop}_{6S}}{\mathcal{B}^{\rm loop}_{7S}} =4.15\sim4.45 \qquad (1.0\leq\alpha\leq3.0).
	\label{eq:BR_loop_ratio}
\end{equation}
For a common value of $\alpha$, the $\eta_c(7S)$ baryonic branching fraction is therefore about $22\%\sim 24$\% of the $\eta_c(6S)$ result.  This ratio is much less sensitive to the correlated cutoff variation than the individual branching fractions.  A direct measurement of one of these pseudoscalar decays to $\Lambda_c^+\bar{\Lambda}_c^-$ would constrain the effective cutoff within the present hadron-loop model.

The hadronic loop calculation includes intermediate ground-state $D^{(*)}$ and $D_s^{(*)}$ mesons.  The channels involving excited charmed mesons $D_J$ dominate several QPC partial widths in Table~\ref{tab:qpc}, but their couplings to ground-state baryons have not been measured and cannot presently be included reliably.  We therefore retain only the ground-state loops and regard the resulting rates as ground-state-loop estimates rather than complete hadron-loop predictions. The omission of the excited channels, together with the uncertainties in the  coupling constants from SU(3)-flavor breaking, is a major source of uncertainty in the absolute decay rates.

\subsection{Naive-factorization estimate of $B^0\to K^0\eta_c(nS)$}
\label{sec:Bdecay}

We next provide a rough estimate of the weak production rates for $B^0\to K^0\eta_c(nS)$ ($n=6,7$).  Following the standard treatment of the color-suppressed $B\to K\eta_c(1S)$ decays~\cite{Song:2002,Cheng:1999}, the naive-factorization amplitude reads

\begin{multline}
\mathcal A_{\rm NF}(B^0\to K^0\eta_c(nS)) =i\frac{G_F}{\sqrt{2}}{V_{cb}^*V_{cs}}a_2^{\rm eff}f_{\eta_c(nS)} \\
\times(m_B^2-m_K^2)F_0^{B\to K}(m_{\eta_c(nS)}^2).
\label{eq:naive-factorization}
\end{multline}
Here, the decay constant is defined through $\langle\eta_c(nS)(q)|\bar c\gamma_\mu\gamma_5c|0\rangle =if_{\eta_c(nS)}q_\mu$, whereas the scalar $B\to K$ form factor satisfies $q^\mu\langle K|\bar s\gamma_\mu b|B\rangle =(m_B^2-m_K^2)F_0^{B\to K}(q^2)$.  These matrix elements yield the factorized amplitude in Eq.~\eqref{eq:naive-factorization}. Upon neglecting CP violation in neutral-kaon mixing, the rate to the reconstructed $K_S^0$ final state is one-half of the corresponding $K^0$ rate.  The two-body width is given by
\begin{equation}
\Gamma(B^0\to K^0\eta_c(nS)) =\frac{|\boldsymbol p_K|}{8\pi m_B^2}|\mathcal A_{\rm NF}|^2.
\label{eq:factorization-width}
\end{equation}
Calculations of highly excited pseudoscalar charmonium states give different values of $f_{\eta_c(6S)}$, depending on the treatment of relativistic and QCD corrections~\cite{Soni:2018,Patel:2009,Kher:2018}.  We therefore use $f_{\eta_c(6S)}=180\sim 250$~MeV as a representative input interval.  We also consider $f_{\eta_c(6S)}=350$~MeV as an optimistic benchmark, which is motivated by lattice, sum-rule, and phenomenological determinations for the ground state~\cite{Davies:2010,Becirevic:2014,Edwards:2001}.  We use this interval as a reference scale for the fitted $6S$ contribution.  Direct calculations of $f_{\eta_c(7S)}$ are not available, so we use $f_{\eta_c(7S)}=150\sim 300$~MeV as an exploratory interval that allows for a decrease relative to the $6S$ state and for the spread among higher radial excitations.

For the kinematics, we take $m_{\eta_c(6S)}=4.629$~GeV and $m_{\eta_c(7S)}=4.719$~GeV, which are the masses fixed for the two charmonium amplitudes in the $H_{67}$ fit.  The central values from the HPQCD parametrization~\cite{Parrott:2022} are $F_0^{B\to K}=0.762$ and 0.810 at these two masses.  The uncertainty of the form factor is not included in the ranges below. We take $a_2^{\rm eff}=0.25$~\cite{Cheng:1999}, together with the PDG values of the CKM elements, masses, and $B^0$ lifetime~\cite{ParticleDataGroup:2026aaa}.  Since $a_2^{\rm eff}$ is process-dependent and has not been calibrated for such  highly excited charmonia, we use this value as the input for the estimates below.  With these inputs, naive factorization gives the following branching fractions: 
\begin{align}
\mathcal B(B^0\to K_S^0\eta_c(6S))_{\rm NF} &=0.78\times10^{-4} \left(\frac{a_2^{\rm eff}}{0.25}\right)^2 \left(\frac{f_{\eta_c(6S)}}{200~\mathrm{MeV}}\right)^2, \nonumber\\
\mathcal B(B^0\to K_S^0\eta_c(7S))_{\rm NF} &=0.55\times10^{-4} \left(\frac{a_2^{\rm eff}}{0.25}\right)^2 \left(\frac{f_{\eta_c(7S)}}{200~\mathrm{MeV}}\right)^2.
\label{eq:factorization-results}
\end{align}
The adopted ranges of the decay constants therefore yield
\begin{equation}
\begin{split}
\mathcal B(B^0\to K_S^0\eta_c(6S))_{\rm NF}&=(0.63\sim 1.22)\times10^{-4},\\
\mathcal B(B^0\to K_S^0\eta_c(7S))_{\rm NF}&=(0.31\sim 1.23)\times10^{-4}.
\end{split}
\label{eq:factorization-ranges}
\end{equation}
The optimistic benchmark $f_{\eta_c(6S)}=350$~MeV raises the $6S$ result to $2.40\times10^{-4}$.  Varying $a_2^{\rm eff}$ from 0.20 to 0.30 changes either rate by a factor of $0.64\sim 1.44$.

\subsection{Extraction of the $\eta_c(6S)$ branching-fraction product and comparison with theoretical estimates}
\label{sec:area}

\begingroup

In the fixed $H_{67}$ fit, the fraction associated with the $\eta_c(6S)$ amplitude in the $m(\Lambda_c^+\bar{\Lambda}_c^-)$ projection is defined relative to the background-subtracted signal as
\begin{equation}
 F_{\eta_c(6S)}= \frac{\sum_i N_i^{\eta_c(6S)}} {\sum_i\left(N_i^{\rm tot}-N_i^{\rm bkg}\right)} ={2.37\%}.
 \label{eq:eta6s-area-fraction}
\end{equation}
Here, $N_i^{\eta_c(6S)}$ is the contribution obtained from the squared $\eta_c(6S)$ amplitude, $N_i^{\rm tot}$ is the total fitted yield, and $N_i^{\rm bkg}$ is the fixed experimental background.  The denominator contains the complete signal intensity, including the interference terms, and the nonresonant contribution is retained as part of the physical three-body decay.  As a cross-check, the fixed $H_6$ solution gives  $F_{\eta_c(6S)}={1.85\%}$, showing that the extracted fraction changes little when the $\eta_c(7S)$ amplitude is omitted.

Using the narrow-width approximation and the LHCb result $\mathcal B(B^0\to K_S^0\Lambda_c^+\bar{\Lambda}_c^-) \simeq2.60\times10^{-4}$, the fraction in Eq.~\eqref{eq:eta6s-area-fraction} gives

\begin{align}
 &\mathcal B(B^0\to K_S^0\eta_c(6S))\times \mathcal B(\eta_c(6S)\to\Lambda_c^+\bar{\Lambda}_c^-)\nonumber\\
 &\qquad\simeq F_{\eta_c(6S)} \mathcal B(B^0\to K_S^0\Lambda_c^+\bar{\Lambda}_c^-) \simeq{6.2\times10^{-6}}.
 \label{eq:effective-chain-rate}
\end{align}

An independent theoretical estimate of the same complete-chain product is obtained by multiplying the naive-factorization result for $\mathcal B(B^0\to K_S^0\eta_c(6S))$ in Eq.~\eqref{eq:factorization-ranges} by the hadron-loop result for $\mathcal B(\eta_c(6S)\to\Lambda_c^+\bar{\Lambda}_c^-)$.  We select $1.6\leq\alpha\leq2.8$ as a reference interval, over which $\mathcal B(\eta_c(6S)\to\Lambda_c^+\bar{\Lambda}_c^-)$ varies from $5.96\times10^{-5}$ to $1.07\times10^{-3}$.  This gives
\begin{equation}
\begin{aligned}
 &\mathcal B(B^0\to K_S^0\eta_c(6S))_{\rm NF}\times \mathcal B(\eta_c(6S)\to\Lambda_c^+\bar{\Lambda}_c^-)_{\rm loop} \\
 &\qquad=(3.8\times10^{-9}\sim 1.3\times10^{-7}).
\end{aligned}
 \label{eq:factorized-chain-rate}
\end{equation}
If the optimistic factorization input $f_{\eta_c(6S)}=350$~MeV is used while $a_2^{\rm eff}=0.25$ is kept fixed, the upper end of the theoretical branching-fraction product becomes $2.6\times10^{-7}$.  Over the main input ranges, the product extracted from the fit is therefore about $1.7\sim 3.2$ orders of magnitude larger than the product of the naive-factorization production branching fraction and the hadron-loop decay branching fraction.  Even with $f_{\eta_c(6S)}=350$~MeV, the product extracted from the fit remains about $1.4$ orders of magnitude larger.

The weak-production calculation is a natural first place to examine this difference.  Besides the uncertainties in $a_2^{\rm eff}$, the decay constant, and the transition form factor, it remains an open question whether the short-distance factorization mechanism provides the dominant contribution to the production of a charmonium state as highly excited as $\eta_c(6S)$. Long-distance rescattering through $D_s^{(*)}\bar D^{(*)}$ intermediate states can bypass the color suppression of the direct process, as considered in analogous $B$-decay mechanisms~\cite{Duan:2021}.  Such contributions are not included here.  A quantitative calculation would require a phenomenological cutoff analogous to that used in hadron-loop models, but no empirically calibrated range of this cutoff is available for the production of highly excited charmonia in $B$ decays.  A controlled numerical prediction of the long-distance production rate is therefore not possible at present.

The hadron-loop prediction for $\eta_c(6S)\to\Lambda_c^+\bar{\Lambda}_c^-$ is also model dependent, as discussed above.  Nevertheless, a branching fraction into the charmed-baryon pair reaching the order of $10^{-3}$ is qualitatively compatible with a highly excited charmonium whose total width is dominated by open-charm meson channels. If the product extracted from the fit is combined instead with the reference hadron-loop range, the weak-production branching fraction inferred from it would be
\begin{equation}
\begin{aligned}
 &\left.\mathcal B(B^0\to K_S^0\eta_c(6S))\right|_{\rm fit/loop} \\
 &\qquad=\left(\frac{{6.2\times10^{-6}}}{1.07\times10^{-3}}\right) \sim\left(\frac{{6.2\times10^{-6}}}{5.96\times10^{-5}}\right)\\
 &\qquad\simeq{(5.8\times10^{-3})\sim(1.0\times10^{-1})},
\end{aligned}
 \label{eq:inferred-production-rate}
\end{equation}
or approximately $0.58\%\sim10.4\%$.  Branching fractions at the $10^{-3}$ level are not in themselves excluded for $B$-meson decays into a kaon plus a charmonium state.  For example, the PDG averages include $\mathcal B(B^+\to K^+\eta_c)\simeq1.10\times10^{-3}$, $\mathcal B(B^+\to K^+J/\psi)\simeq1.02\times10^{-3}$ for ground-state charmonia~\cite{ParticleDataGroup:2026aaa}.  The lower end of Eq.~\eqref{eq:inferred-production-rate}, $5.8\times10^{-3}$, is already about the same order of magnitude as these measured rates and about 50 times the upper end of the main naive-factorization range.  More importantly, the upper end of 10.4\% is clearly not physically reasonable for an exclusive color-suppressed $B$-meson decay into a kaon plus charmonium.  This unphysical upper value indicates that the hadron-loop calculation of $\eta_c(6S)\to\Lambda_c^+\bar{\Lambda}_c^-$ should favor cutoff $\alpha\sim3$ or larger.

Since open-charm modes are expected to dominate the decays of highly excited charmonia, measurements of $B$ decays into a kaon and different open-charm meson pairs, such as $B\to K D^{(*)}\bar D^{(*)}$, are particularly important.  Such measurements would serve two purposes. First, a common structure near 4.63~GeV in several charmed-meson-pair spectra would provide an independent test of the possible $\eta_c(6S)$ signal.  Second, measurements of the corresponding exclusive rates, particularly for the dominant open-charm decay modes, would also test whether the $B\to K\eta_c(6S)$ production branching fraction can reach the scale implied by the present fit and would provide empirical input for assessing the relative importance of short- and long-distance production mechanisms.

\endgroup

\section{Summary}
\label{sec:summary}

\begingroup

In this work, we have focused on the event accumulation near 4.63~GeV in the $\Lambda_c^+\bar{\Lambda}_c^-$ spectrum of $B^0\to K_S^0\Lambda_c^+\bar{\Lambda}_c^-$, and have suggested that this enhancement structure may originate from the contribution of the missing charmonium $\eta_c(6S)$ state.  Starting with a baseline model comprising the nonresonant amplitude and the known $\Xi_c(2923)$ and $\Xi_c(2939)$ states, we have tested the intermediate resonance contributions of the highly excited charmonia $\eta_c(6S)$ and $\eta_c(7S)$ in the $B^0\to K_S^0\Lambda_c^+\bar{\Lambda}_c^-$ process, using fixed representative masses and widths selected from the spectroscopic ranges.

The modified GI potential model predicts an $\eta_c(6S)$ mass of $4581.4\sim 4630.7$~MeV, covering the possible structure near 4.63~GeV in the $\Lambda_c^+\bar\Lambda_c^-$ spectrum.  The $\eta_c(6S)$ is therefore a promising candidate for explaining this potential enhancement.  With the resonance parameters of $\eta_c(6S)$ fixed at $(M_1,\Gamma_1)=(4629.00,24.20)$~MeV, including this contribution improves the description of the enhancement structure in the $\Lambda_c^+\bar\Lambda_c^-$ spectrum.  The additional $\eta_c(7S)$ contribution, with $(M_2,\Gamma_2)=(4718.84,22.84)$~MeV, further improves the description of the higher-mass region.

The fixed $H_{67}$ fit yields a background-subtracted $\eta_c(6S)$ fraction of 2.37\%.  Using the narrow-width approximation, this fraction gives a complete-chain branching-fraction product of about $6.2\times10^{-6}$. We have also calculated the hadron-loop branching fraction for $\eta_c(6S)\to\Lambda_c^+\bar{\Lambda}_c^-$.  Independently, the product of the naive-factorization production branching fraction and the hadron-loop decay branching fraction $\mathcal B(B^0\to K_S^0\eta_c(6S))_{\rm NF}\times \mathcal B(\eta_c(6S)\to\Lambda_c^+\bar{\Lambda}_c^-)_{\rm loop}$ lies about $1.7\sim 3.2$ orders of magnitude below this value extracted from the fit.  This difference may reflect limitations of naive factorization and important contributions from long-distance mechanisms for estimating  the production of such a highly excited charmonium state in the $B$-meson decay.

The possible $\eta_c(6S)$ signal provides a new opportunity to search for highly excited pseudoscalar charmonia in $B$-meson decays.  More precise measurements of $B^0\to K_S^0\Lambda_c^+\bar{\Lambda}_c^-$, together with searches for a structure near 4.63~GeV in $B$ decays into a kaon and different open-charm meson pairs, will be important for establishing whether this enhancement is associated with the missing charmonium $\eta_c(6S)$.  Such measurements would also test the decay pattern of the $\eta_c(6S)$ candidate and clarify the production behavior of highly excited charmonia in $B$-meson decays. These measurements provide valuable targets for Belle II and LHCb.
\endgroup

\vfil

\begin{acknowledgments}
This work is supported by the Natural Science Foundation of Gansu Province (Nos. 26RCKA012 and 25JRRA799), the National Natural Science Foundation of China under Grants No. 12335001, No. 12247101, No. 12405088 and No. 12547101, the ``111 Center'' under Grant No. B20063, the Fundamental Research Funds for the Central Universities (lzujbky-2023-stlt01), and Lanzhou City High-Level Talent Funding. 
\end{acknowledgments}
\appendix

\onecolumngrid
\section{Triangle-loop amplitudes for $\eta_c(nS)\to\Lambda_c^+\bar\Lambda_c^-$}
\label{app:etac}

The hadron-loop amplitude for $\eta_c(nS)\to\Lambda_c^+\bar\Lambda_c^-$ is given by Eq.~\eqref{eq:hadron_loop}.  We use $p_1$ for the incoming $\eta_c$ momentum and $p_2$ ($p_3$) for the outgoing $\Lambda_c$ ($\bar\Lambda_c$) momentum.  The loop momenta $q_1$ and $q_2$ follow the orientation in Fig.~\ref{fig:loop_LcLc}; the common Feynman $i\epsilon$ prescription and overall factors of $i$ are understood in the convention of Eq.~\eqref{eq:hadron_loop}. The vertex products $\mathcal{M}\equiv V_{M_1M_2}\,T_{M_1B}\,T_{M_2B}$ appearing in the numerator, for the six diagrams in Fig.~\ref{fig:loop_LcLc}, are listed below.  The propagator denominators $(q_1^2-m_{M_1}^2)$, $(q_2^2-m_{M_2}^2)$, $(q^2-m_B^2)$ and the form factor $\mathcal{F}^2(q^2)$ are given explicitly in Eq.~\eqref{eq:hadron_loop}.  The momenta satisfy $p_1=q_1+q_2$ and $q=q_2-p_2=p_3-q_1$. 

For the $D^{(*)}\bar D^{(*)}$ intermediate states with $N$ exchange [diagrams (a--c) in Fig.~\ref{fig:loop_LcLc}]:
\begin{align}
\mathcal{M}_a &= 2\,g_{\eta_c D D^*}\,(q_{2\alpha}-q_{1\alpha})\, \Pi^{\alpha\beta}(q_1;m_{D^*}) \,\bar u(p_2)\Bigl[ i g_{D^*\Lambda_c N}\,\gamma_\beta + i\frac{\kappa_{D^*\Lambda_c N}}{m_{\Lambda_c}+m_N}\,q_{1k}\,\sigma^{k}{}_\beta\Bigr] ({\slashed{q}}+m_N) \bigl(i g_{D\Lambda_c N}\,\gamma_5\bigr) v(p_3), \\[4pt]
\mathcal{M}_b &= 2\,g_{\eta_c D D^*}\,(q_{2\lambda}-q_{1\lambda})\, \Pi^{\lambda\rho}(q_2;m_{D^*})\,\bar u(p_2)\, \bigl(i g_{D\Lambda_c N}\,\gamma_5\bigr) ({\slashed{q}}+m_N) \Bigl[ i g_{D^*\Lambda_c N}\,\gamma_\rho + i\frac{\kappa_{D^*\Lambda_c N}}{m_{\Lambda_c}+m_N}\,q_{2l}\,\sigma^{l}{}_\rho\Bigr] v(p_3), \\[4pt]
\mathcal{M}_c &= -2i\,g_{\eta_c D^* D^*}\,p_{1n}\,(q_{1m}-q_{2m})\, \varepsilon^{\alpha\lambda mn} \,\Pi_{\alpha}{}^{\beta}(q_1;m_{D^*})\, \Pi_{\lambda}{}^{\rho}(q_2;m_{D^*})\,\bar u(p_2) \nonumber\\
&\quad\times\Bigl[ i g_{D^*\Lambda_c N}\,\gamma_\beta + i\frac{\kappa_{D^*\Lambda_c N}}{m_{\Lambda_c}+m_N}\,q_{1k}\,\sigma^{k}{}_\beta\Bigr] ({\slashed{q}}+m_N) \Bigl[ i g_{D^*\Lambda_c N}\,\gamma_\rho + i\frac{\kappa_{D^*\Lambda_c N}}{m_{\Lambda_c}+m_N}\,q_{2l}\,\sigma^{l}{}_\rho\Bigr] v(p_3).
\end{align}

For the $D_s^{(*)}\bar D_s^{(*)}$ intermediate states with $\Lambda$ exchange [diagrams (d--f)]:
\begin{align}
\mathcal{M}_d &= 2\,g_{\eta_c D_s D_s^*}\,(q_{2\alpha}-q_{1\alpha})\, \Pi^{\alpha\beta}(q_1;m_{D_s^*}) \,\bar u(p_2)\Bigl[ i g_{D_s^*\Lambda_c\Lambda}\,\gamma_\beta + i\frac{\kappa_{D_s^*\Lambda_c\Lambda}}{m_{\Lambda_c}+m_\Lambda}\,q_{1k}\,\sigma^{k}{}_\beta\Bigr] ({\slashed{q}}+m_\Lambda) \bigl(i g_{D_s\Lambda_c\Lambda}\,\gamma_5\bigr) v(p_3), \\[4pt]
\mathcal{M}_e &= 2\,g_{\eta_c D_s D_s^*}\,(q_{2\lambda}-q_{1\lambda})\, \Pi^{\lambda\rho}(q_2;m_{D_s^*}) \,\bar u(p_2)\, \bigl(i g_{D_s\Lambda_c\Lambda}\,\gamma_5\bigr) ({\slashed{q}}+m_\Lambda) \Bigl[ i g_{D_s^*\Lambda_c\Lambda}\,\gamma_\rho + i\frac{\kappa_{D_s^*\Lambda_c\Lambda}}{m_{\Lambda_c}+m_\Lambda}\,q_{2l}\,\sigma^{l}{}_\rho\Bigr] v(p_3), \\[4pt]
\mathcal{M}_f &= -2i\,g_{\eta_c D_s^* D_s^*}\,p_{1n}\,(q_{1m}-q_{2m})\, \varepsilon^{\alpha\lambda mn} \,\Pi_{\alpha}{}^{\beta}(q_1;m_{D_s^*})\, \Pi_{\lambda}{}^{\rho}(q_2;m_{D_s^*}) \,\bar u(p_2) \nonumber\\
&\quad\times\Bigl[ i g_{D_s^*\Lambda_c\Lambda}\,\gamma_\beta + i\frac{\kappa_{D_s^*\Lambda_c\Lambda}}{m_{\Lambda_c}+m_\Lambda}\,q_{1k}\,\sigma^{k}{}_\beta\Bigr] ({\slashed{q}}+m_\Lambda) \Bigl[ i g_{D_s^*\Lambda_c\Lambda}\,\gamma_\rho + i\frac{\kappa_{D_s^*\Lambda_c\Lambda}}{m_{\Lambda_c}+m_\Lambda}\,q_{2l}\,\sigma^{l}{}_\rho\Bigr] v(p_3).
\end{align}
Here, $\Pi^{\mu\nu}(q;m)=q^\mu q^\nu/m^2-g^{\mu\nu}$ is the spin-1 propagator tensor numerator; the pseudoscalar $D$ and $D_s$ legs contribute a factor of unity. The $V_{M_1M_2}$ vertices follow from Eqs.~\eqref{eq:L_DDstar} and \eqref{eq:L_DstarDstar}, and the $T_{MB}$ vertices from Eq.~\eqref{eq:L_Dstar}.

\twocolumngrid

\bibliography{ref}

\end{document}